\documentclass[twocolumn,showpacs,preprintnumbers,amsmath,amssymb]{revtex4}

\usepackage{graphicx}
\usepackage{dcolumn}
\usepackage{bm}
\usepackage{epsfig}

\begin{document}


\title{Neutron scattering evidence on the nature of the boson peak}

\author{U. Buchenau}
 \email{u.buchenau@fz-juelich.de}
\author{A.Wischnewski}
\author{M. Ohl}
\affiliation{%
Institut f\"ur Festk\"orperforschung, Forschungszentrum J\"ulich\\
Postfach 1913, D--52425 J\"ulich, Federal Republic of Germany
}%
\author{E. Fabiani}
\affiliation{%
Institute Laue-Langevin, BP 156, F-38042 Grenoble Cedex 9, France
}%

\date{July 5, 2004}

\begin{abstract}
A close inspection of neutron spectra of glass formers in the
frequency region above the boson peak does not show the constant
eigenvalue density expected for a random dynamical matrix, but a
slow decrease towards higher eigenvalues. In 1,4-polybutadiene and
selenium, the slope becomes steeper with increasing temperature in
the undercooled liquid, suggesting an influence of the vibrational
entropy. One can describe the behaviour quantitatively in terms of
a balance between level repulsion (from the randomness of the
dynamical matrix) and vibrational entropy. On the basis of this
evidence, the boson peak must be a crossover from such a balance
at higher frequencies to a mixture of sound waves and additional
excitations at low frequency.
\end{abstract}

\maketitle

There is as yet no generally accepted explanation of the boson
peak in the neutron or Raman spectrum of glasses
\cite{gurevich,schirmacher,elliott,nakayama,gotze,sokolov,parisi,ruocco}.
This is a broad peak at an energy transfer of a few meV, where
simple crystals have only sound waves. Glasses seem to have a
sizeable amount of excess vibrations at this boson peak. At
present, it is not clear which driving force brings these
vibrations down into the low-frequency region.

In this paper, we present for the first time experimental evidence
for a subtle but important driving force neglected so far, the
vibrational entropy.

The glass freezes at a relatively high temperature (the glass
temperature $T_g$). At this temperature, the thermal energy is an
order of magnitude higher than the vibrational energy quantum at
the boson peak. Therefore any theoretical consideration of the
origin of the boson peak should take the vibrational entropy into
account, because it tends to bring the eigenmodes down to low
frequencies. A lower frequency allows to distribute the thermal
energy between more energy levels and thus increases the entropy
\cite{barron,hui}.

In order to see this influence in experimental data, one plots the
eigenvalue density $p(\lambda)$ against the eigenvalue $\lambda$
on a logarithmic scale. Since the neutron measurement supplies the
frequency as energy transfer $E=\hbar\omega$, we define the
eigenvalue $\lambda=E^2$ and measure it in $meV^2$. The eigenvalue
density $p(\lambda)=g(E)/2E$, where $g(E)$ is the conventional
vibrational density of states. The boson peak is a broad peak in
$g(E)/E^2$. Therefore it is more like a shoulder in $p(\lambda)$.

All measurements presented were done on the time-of-flight
spectrometer IN6 at the High Flux Reactor of the Institut
Laue-Langevin in Grenoble, France. The wavelength of the incoming
neutrons was $4.1\ \AA$. Some data have already been published,
but not in the scaling presented here. The samples had a
scattering probability of about 10 \% to provide a reasonable
balance between signal strength and multiple scattering
contamination.

The data were evaluated in a new scheme for the determination of
the vibrational density of states from coherent and incoherent
inelastic neutron scattering data, developed by two of us at the
Institut Laue-Langevin at Grenoble \cite{fabiani}.

\begin{figure}[b]
\hspace{-0cm} \vspace{0cm} \epsfig{file=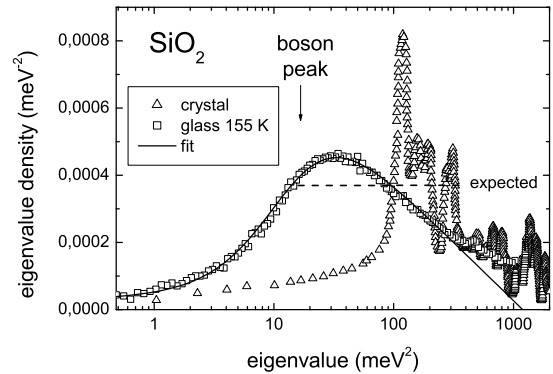,width=8
cm,angle=0} \vspace{0cm} \caption{Eigenvalue density in
crystalline $\alpha$-quartz \cite{strauch} and in vitreous silica
\cite{bu1}. The line is a fit in terms of eq. (\ref{intpol}).}
\end{figure}

Fig. 1 shows the classical case of silica, the glass spectrum of a
measurement at 155 K \cite{bu1} and the crystal spectrum from a
lattice dynamical calculation \cite{strauch}. In the region around
10 $meV$, the crystal shows pronounced van-Hove singularities. The
disorder should broaden these low-lying van-Hove singularities
into a quasi-constant eigenvalue density on the basis of Wigner's
famous solution \cite{wig} of the random-matrix problem. Instead,
we observe a slow decrease towards higher eigenvalues, linear on
this logarithmic scale over nearly a decade. Our results are in
qualitative agreement with recent nuclear inelastic scattering
data \cite{chumakov}, which find a stronger than $1/E$-dependence
in $g(E)/E^2$ above the boson peak.

Let us see whether we can justify such a linear decrease on the
basis of the concept of vibrational entropy. The tendency to form
a constant eigenvalue density from the randomness of the dynamical
matrix can be described in terms of an energetic level repulsion
term \cite{schirmacher,wig}. We assume this level repulsion term
to be proportional to the eigenvalue density, with a
proportionality factor $B$. The free energy $F_\lambda$ of an
eigenmode at the eigenvalue $\lambda$ is then
\begin{equation}\label{free}
  F_\lambda=Bp(\lambda)+\frac{k_BT}{2}\ln\lambda.
\end{equation}
The second term is the vibrational entropy \cite{barron,hui} in
the classical limit $E<<k_BT$.

From eq. (\ref{free}) we obtain the eigenvalue density
\begin{equation}\label{dens}
p(\lambda)=\frac{F_\lambda}{B}-\frac{k_BT}{2B}\ln\lambda,
\end{equation}
which is the linear relation to $\ln\lambda$ we are looking for.

\begin{figure}[b]
\hspace{-0cm} \vspace{0cm} \epsfig{file=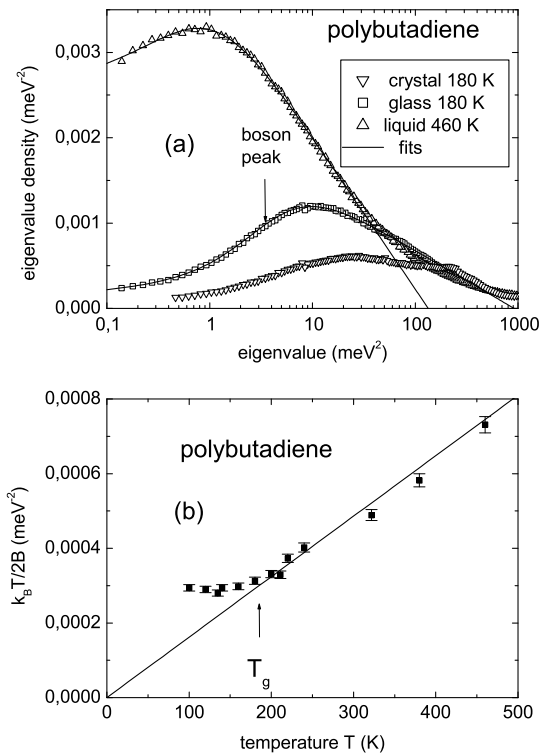,width=8
cm,angle=0} \vspace{0cm} \caption{(a) Eigenvalue density in
crystalline, glassy and liquid polybutadiene \cite{zorn}. Lines
are fits in terms of eq. (\ref{intpol}). (b) Temperature
dependence of the fitted slope $k_BT/2B$ in polybutadiene.}
\end{figure}

The slope $-k_BT/2B$ should be proportional to the temperature, at
least as long as the eigenvalue density is able to adapt itself to
a new thermodynamic equilibrium.  This can be used to check the
validity of the scheme in the undercooled liquid.

Fig. 2 (a) demonstrates the strong temperature change of the
eigenvalue density in the heavily studied example of
1,4-polybutadiene \cite{zorn}. The temperature dependence of the
fitted slope $k_BT/2B$ in Fig. 2 (b) follows the expectation of
eq. (\ref{dens}) over a wide temperature range, from high
temperatures in the liquid down until the sample freezes at the
glass temperature.

\begin{figure}[b]
\hspace{-0cm} \vspace{0cm} \epsfig{file=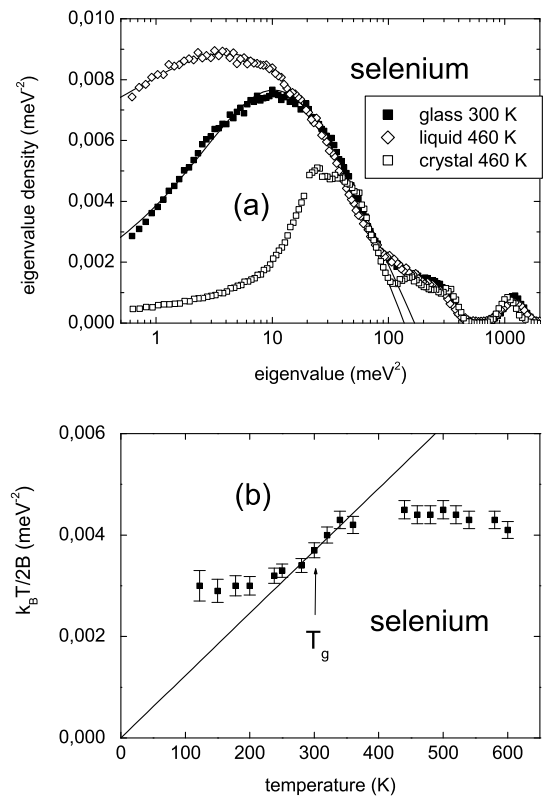,width=8
cm,angle=0} \vspace{0cm} \caption{(a) Eigenvalue density in
crystalline, glassy and liquid selenium \cite{andrew}. Lines are
fits in terms of eq. (\ref{intpol}). (b) Temperature dependence of
the fitted slope $k_BT/2B$ in selenium.}
\end{figure}

Polybutadiene is a clearcut example for the validity of eq.
(\ref{dens}) over a wide temperature range, with a level repulsion
constant $B$ which does not depend on temperature. This is not
always the case. One finds a slight temperature dependence of $B$
in the third example, selenium, as shown in Fig. 3. These are
already published data \cite{andrew}, complemented by new
experiments with a thinner sample. In this case, the
proportionality of the fitted slope to temperature is limited to a
small temperature range around $T_g$, extending to slightly lower
values as $T_g$. But within that range, the selenium data again
support the physical picture of a balance between vibrational
entropy and level repulsion at frequencies above the boson peak.

If this picture is true, the boson peak must be a crossover
phenomenon between such a balance at higher frequency and a quite
different physical reality at low frequency. We know from many
experiments that we have well-defined sound waves at frequencies
below the boson peak. The question is: What is the physical
mechanism which causes such a crossover? Is it the anharmonicity
which reduces the number of levels of a given eigenmode potential
as one approaches the eigenvalue zero? Or is it the interaction
between resonant modes and sound waves?

Our experimental evidence does not provide a direct answer to this
central question.  But it suggests to look for a reasonable
interpolation scheme between low and high frequency. At low
frequency, one has again a reliable theoretical picture, namely an
elastic medium with well-defined sound waves coexisting with a
small number of additional modes. There, the neutron spectra
should be well described by the soft-potential expression
\begin{equation}\label{plow}
p_{low}(\lambda)=\frac{3}{2}\frac{\lambda^{1/2}}{\omega_D^3}
+\frac{1}{2}f_{vib}\lambda^{3/2}+\frac{1}{2}f_{rel}.
\end{equation}
This expression for the low-frequency spectrum is derived from the
soft-potential model \cite{parshin}, an extension of the tunneling
model of the low temperature anomalies of glasses. The first term
contains the Debye frequency $\omega_D$ and describes the sound
waves. In addition to the sound waves, the model postulates a
continuous distribution of additional modes around the eigenvalue
zero. The positive eigenvalues provide vibrational resonant modes
coexisting with the sound waves, the second term of eq.
(\ref{plow}). The negative eigenvalues, in principle unstable
modes, are supposed to be stabilized by the anharmonic fourth
order term of the mode potential. They lead to double-well
potentials, giving rise to tunneling states at low temperatures
and to classical relaxation at higher temperatures. The third term
of eq. (\ref{plow}) is the soft-potential expectation for this
classical relaxation spectrum. $f_{vib}$ and $f_{rel}$ are given
in terms of the parameters of the soft-potential model
\cite{ramos}
\begin{equation}\label{fvib}
f_{vib}=\frac{1}{24}\frac{P_sM}{\rho}\left(\frac{\hbar}{W}\right)^5.
\end{equation}
and
\begin{equation}\label{frel}
f_{rel}\approx\frac{1}{2}\frac{P_sM}{\rho}\left(\frac{\hbar}{W}\right)^2
\left(\frac{k_BT}{W}\right)^{3/4}.
\end{equation}
Here $P_s$ is the density of additional modes around the
eigenvalue zero, $M$ is the average atomic mass, $\rho$ is the
mass density and $W$ is the crossover energy between vibrational
and tunneling states at low temperatures.

The soft-potential model has been checked against the
low-temperature glass anomalies in the specific heat, the thermal
conductivity and in the mechanical loss \cite{parshin,ramos}. The
model predictions were found to be essentially correct, with an
important exception: as soon as the barrier height of the
double-well potentials begins to be a sizeable fraction of the
thermal energy at the glass transition, the measured classical
relaxation gets much weaker than the soft-potential prediction
\cite{kasper,ramos}. As we will see, the same effect appears in
the neutron spectra at elevated temperature.

In order to check the neutron data, one needs a suitable
interpolation scheme between eq. (\ref{plow}) below the boson peak
and eq. (\ref{dens}) above. As it turns out, it is convenient to
use
\begin{equation}\label{intpol}
p(\lambda)=\frac{1}{1/p_{low}(\lambda)+1/p_{high}(\lambda)},
\end{equation}
where $p_{high}(\lambda)$ is given by eq. (\ref{dens}).

Eq. (\ref{intpol}) is an elegant way of interpolating from one
eigenvalue density to the other, without introducing an additional
crossover parameter. One can characterize the whole spectrum with
five reasonable parameters, each of which has a well-defined
physical meaning either in the low-frequency or in the
high-frequency range.

\begin{figure}[b]
\hspace{-0cm} \vspace{0cm} \epsfig{file=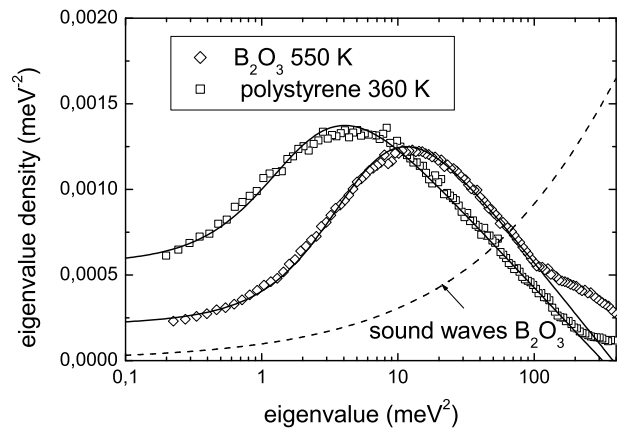,width=9
cm,angle=0} \vspace{0cm} \caption{Eigenvalue density in $B_2O_3$
\cite{engberg} and in polystyrene \cite{pecha} at the glass
transition. Lines are fits in terms of eq. (\ref{intpol}).}
\end{figure}

The above-mentioned failure of the soft-potential model is most
clearly seen in the neighborhood of the glass temperature. Fig. 4
shows the neutron eigenvalue spectrum of $B_2O_3$ at its glass
temperature of 550 K \cite{engberg}, together with the sound wave
expectation on the basis of Brillouin scattering measurements at
the same temperature \cite{grims}. There is a sizeable
relaxational signal at the low-frequency end, which allows to fit
$f_{rel}$ with an accuracy of 25 \%. The vibrational contribution
$f_{vib}$ is fitted with the same accuracy. If one compares the
ratio $f_{rel}/f_{vib}$ with the soft-potential expectation of
eqs. (\ref{fvib}) and (\ref{frel}), it is a factor of five too low
(the comparison requires the crossover energy $W$ between
vibrational and tunneling states, which was taken from ref.
\cite{ramos}). The same is true for polystyrene at its glass
temperature 360 K \cite{pecha} in Fig. 4 and for the case of
1,4-polybutadiene at its glass temperature of 180 K in Fig. 2 (a).

While these results cannot be understood in terms of the
soft-potential model, they support independently our conclusion on
the importance of the vibrational entropy. $f_{vib}$ describes the
low-frequency additional vibrations at low energy transfer $E$.
These are eigenmodes which still have a positive eigenvalue.
Therefore they profit in full from the vibrational entropy, having
an entropy contribution to the free energy of $k_BT\ln(E/k_BT)$.
This is different for $f_{rel}$, which stems from eigenmodes with
a negative eigenvalue. These modes correspond to relaxational
jumps between neighbouring energy minima, with an entropy
contribution to the free energy of the order $-k_BT\ln 2$. Thus
one expects a Boltzmann factor ratio of the order of $k_BT_g/2E$
between positive and negative eigenvalues close to zero, if both
freeze under otherwise identical conditions at the glass
temperature. Since the thermal energy at the glass transition is
an order of magnitude larger than the energy transfer at the boson
peak, this is the observed factor of five.

To summarize, we find threefold experimental evidence for the
influence of vibrational entropy on the neutron spectra of glass
formers. The first is a decrease of the eigenvalue density above
the boson peak, linear in a plot against the logarithm of the
eigenvalue (this has already been seen in nuclear inelastic
scattering \cite{chumakov}, though fitted differently). The second
is the temperature dependence of the corresponding slope above the
glass temperature, proportional to temperature in
1,4-polybutadiene and selenium. The third is the weakness of the
relaxational part of the spectrum close to the glass transition,
which is a factor five weaker than the soft-potential expectation
of an equal footing of positive and negative eigenvalues close to
the eigenvalue zero.

These findings do not resolve the boson peak problem, but they set
the stage for a proper theoretical treatment. They show the
existence of a theoretically well-defined region at frequencies
above the boson peak, where crystals have their first van-Hove
singularity. At the glass transition, this region is dominated by
the balance between level repulsion and vibrational entropy; sound
waves are heavily overdamped there \cite{sette}. On the other
hand, the sound wave picture is the correct starting point in a
glass below the boson peak frequency, where the glass is an
elastic medium with a small number of additional excitations. The
boson peak must be the crossover between these two regions.

\end{document}